\documentclass[apl,twocolumn,superscriptaddress]{revtex4-1}
\usepackage{graphicx}
\usepackage{color}
\usepackage{amsfonts}
\usepackage{ulem}

\definecolor{brown}{RGB}{200,100,0}

\def\kbar{$\bar{\mathrm{K}}$}
\def\mbar{$\bar{\mathrm{M}}$}

\def\gbar{$\bar{\mathrm{\Gamma}}$}

\def\WS2{WS$_2$}
\def\MoS2{MoS$_2$}{

\begin{document}
\title{Imaging microscopic electronic contrasts at the interface of single-layer WS$_2$ with oxide and boron nitride substrates}
\author{ S{\o}ren Ulstrup}
\email{address correspondence to ulstrup@phys.au.dk}
\affiliation{Department of Physics and Astronomy, Aarhus University, 8000 Aarhus C, Denmark}
\author{ Roland J. Koch}
\affiliation{ Advanced Light Source, E. O. Lawrence Berkeley National Laboratory, Berkeley, California 94720, USA}
\author{Daniel Schwarz}
\affiliation{ Advanced Light Source, E. O. Lawrence Berkeley National Laboratory, Berkeley, California 94720, USA}
\author{ Kathleen M. McCreary}
\affiliation{Naval Research laboratory, Washington, D.C. 20375, USA}
\author{ Berend T. Jonker}
\affiliation{Naval Research laboratory, Washington, D.C. 20375, USA}
\author{ Simranjeet Singh}
\affiliation{Department of Physics, Carnegie Mellon University, Pittsburgh, Pennsylvania 15213, USA}
\author{ Aaron Bostwick}
\affiliation{ Advanced Light Source, E. O. Lawrence Berkeley National Laboratory, Berkeley, California 94720, USA}
\author{ Eli Rotenberg}
\affiliation{ Advanced Light Source, E. O. Lawrence Berkeley National Laboratory, Berkeley, California 94720, USA}
\author{ Chris Jozwiak}
\affiliation{ Advanced Light Source, E. O. Lawrence Berkeley National Laboratory, Berkeley, California 94720, USA} 
\author{ Jyoti Katoch}
\email{address correspondence to jkatoch@andrew.cmu.edu}
\affiliation{Department of Physics, Carnegie Mellon University, Pittsburgh, Pennsylvania 15213, USA}

\begin{abstract}
The electronic properties of devices based on two-dimensional materials are significantly influenced by interactions with substrate and electrode materials. Here, we use photoemission electron microscopy to investigate the real- and momentum-space electronic structures of electrically contacted single-layer WS$_2$ stacked on hBN, SiO$_2$ and TiO$_2$ substrates. Using work function and X-ray absorption imaging we single-out clean microscopic regions of each interface type and collect the valence band dispersion. We infer the alignments of the electronic band gaps and electron affinities from the measured valence band offsets of WS$_2$ and the three substrate materials using a simple electron affinity rule and discuss the implications for vertical band structure engineering using mixed three- and two-dimensional materials.
\end{abstract}

\maketitle

Semiconducting transition metal dichalcogenides (TMDs) at the single-layer (SL) limit offer entirely new possibilities for fabricating field-effect transistors with atomically thin gating materials and sophisticated contact electrode geometries leading to nanoscale engineered unipolar and ambipolar charge carrier transport \cite{radisavljevic2011,wang2012,jariwala2013,jariwala2014,Gong:2014,Allain2015}. These properties are determined by the electronic band alignments at the vertically stacked interfaces of the active device components, which can be tailored using junctions of TMDs in combination with other TMDs \cite{Schlaf:1999}, TMDs and oxides \cite{McDonnell:2014,Ulstrup:2016} as well as mixed two-dimensional (2D) and three-dimensional (3D) materials \cite{Jariwala:2016}. Understanding how key band alignment parameters such as the valence band (VB) offsets, quasiparticle band gap energies $E_g$, and electron affinities $\chi$, depend on the interface type and quality as well as environmental screening remains an important issue for band structure engineering utilizing 2D materials \cite{Guo:2016}. 

The interplay of these parameters on the electronic properties of SL TMD devices is ideally investigated using spectromicroscopic probes of the electronic structure \cite{Klein2012}. Photoemission electron microscopy (PEEM) is a powerful method in this regard because it offers fast switching between real space and $k$-space imaging modes with work function, core level absorption and VB contrasts \cite{Fujikawa2009,Ulstrup:2016,Koch:2018}. The use of $k$-resolved PEEM for performing microscale angle-resolved photoemission spectroscopy (microARPES) has been an essential tool for observing band structures of SL and few-layer MoS$_2$ \cite{Jin2013,Jin:2015} and WSe$_2$ \cite{Yeh:2015} exfoliated on SiO$_2$ substrates. 

\begin{figure}
\begin{center}
\includegraphics[width=0.49\textwidth]{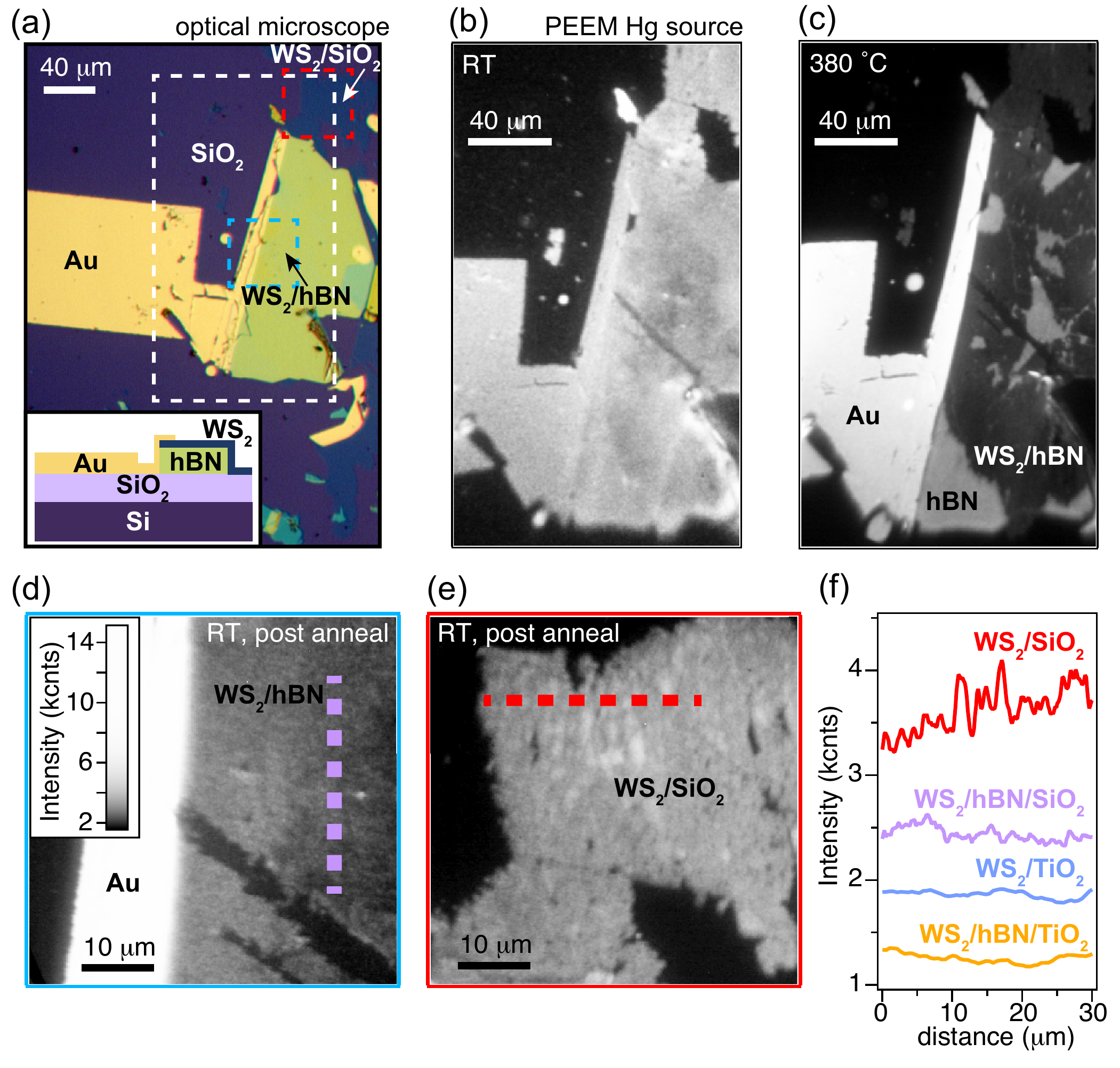}
\caption{Work function contrasts: (a) Optical microscope image of a SiO$_2$ supported sample assembled according to the diagram in the insert. (b)-(c) PEEM images at (b) room temperature before annealing and (c) 380 $^{\circ}$C measured within the dashed white box in (a). (d)-(e) Post annealing PEEM images at room temperature focusing on (d) a WS$_2$/hBN region near the Au contact (blue dashed square in (a)) and (e) a WS$_2$/SiO$_2$ region adjacent to the hBN flake (red dashed square in (a)). (f) Line profiles obtained along the dashed purple and red lines in (d) and (e) and the dashed blue and orange lines in Fig. \ref{fig:2}(b). The color scale in (d) applies to all PEEM images.}
\label{fig:1}
\end{center}
\end{figure}

Here, we use the SPECS PEEM P90 microscope installed at the Microscopic And Electronic STRucture Observatory (MAESTRO) at the Advanced Light Source to investigate the electronic properties of vertical stacks based on SL WS$_2$ transferred on oxide and hexagonal boron nitride (hBN) substrates. The thickness of WS$_2$ is  checked before and after transfer using photoluminescence and Raman spectroscopy as shown in our earlier works \cite{McCreary:2016,Ulstrup:2016}. The influence of the dielectric environment on the electronic properties of SL WS$_2$ is studied using insulating 300 nm SiO$_2$ on Si (SiO$_2$/Si) with relative permittivity $\epsilon_{\mathrm{SiO}_2} = 3.9$ and 0.5~wt~\% Nb-doped rutile TiO$_2$(100) (Shinkosha Co., Ltd) with $\epsilon_{\mathrm{TiO}_2} = 113$ as supporting substrate. We assemble WS$_2$/hBN heterostructures ($\epsilon_{\mathrm{hBN}} \approx 4$) on both oxides utilizing a similar transfer technique as previously reported \cite{Ulstrup:2016,Katoch:2018} and as described further in the Supplementary Material. On SiO$_2$ we deposit an Au electrode that is contacted to both SL WS$_2$ and hBN on the side (see optical microscope image in Fig. \ref{fig:1}(a)) which is essential to avoid charging during photoemission measurements. The Nb doping of TiO$_2$ is sufficient to prevent charging. By shorting the WS$_2$ flake on hBN to the TiO$_2$ we avoid using a metal electrode in this system.

The rationale of using SiO$_2$, hBN and TiO$_2$ as substrates for SL WS$_2$ is three-fold: (i) These materials are commonly used in devices where they are known to exhibit strong variations in interfacial quality with other 2D materials \cite{Chen:2008, Dean:2010,Ulstrup:2016}, (ii) the dielectric properties vary strongly across the interfaces, potentially affecting the electronic bandstructure of the adjacent SL WS$_2$ \cite{Rosner:2016,Raja:2017}, and (iii) the quasiparticle band gaps and electron affinities are very different and thus give rise to substantially different band alignments. Here, we address these key points by first presenting PEEM measurements of electronic contrasts to identify the three types of interfaces and investigate their quality from a photoemission perspective. We then discuss $k$-resolved electronic structure measurements and use these to infer the band alignments of the systems.

The photoemission intensity variations during \textit{in situ} annealing of the SiO$_2$ supported sample to 380~$^{\circ}$C are studied in PEEM using a Hg excitation source as shown in  Figs. \ref{fig:1}(b)-(c). The average contrast levels for Au, SL WS$_2$ and hBN areas are similar before annealing (panel (b)) making it difficult to distinguish the materials. During annealing the intensity of the Au electrode increases (panel (c)). This behavior indicates a lowering of the Au work function giving rise to higher secondary electron emission and therefore higher intensity. The reduction of secondary electron emission from WS$_2$ on hBN during annealing indicates an increase in work function, possibly due to a change in doping caused by the desorption of water. The intensity levels from patches of WS$_2$ on SiO$_2$ and on hBN adjust slightly after cooling down. Most importantly, we observe that there is no sign of Au diffusion on the surface at these annealing conditions in ultra-high vacuum (UHV) at 380 $^{\circ}$C. 

\begin{figure}
\begin{center}
\includegraphics[width=0.4\textwidth]{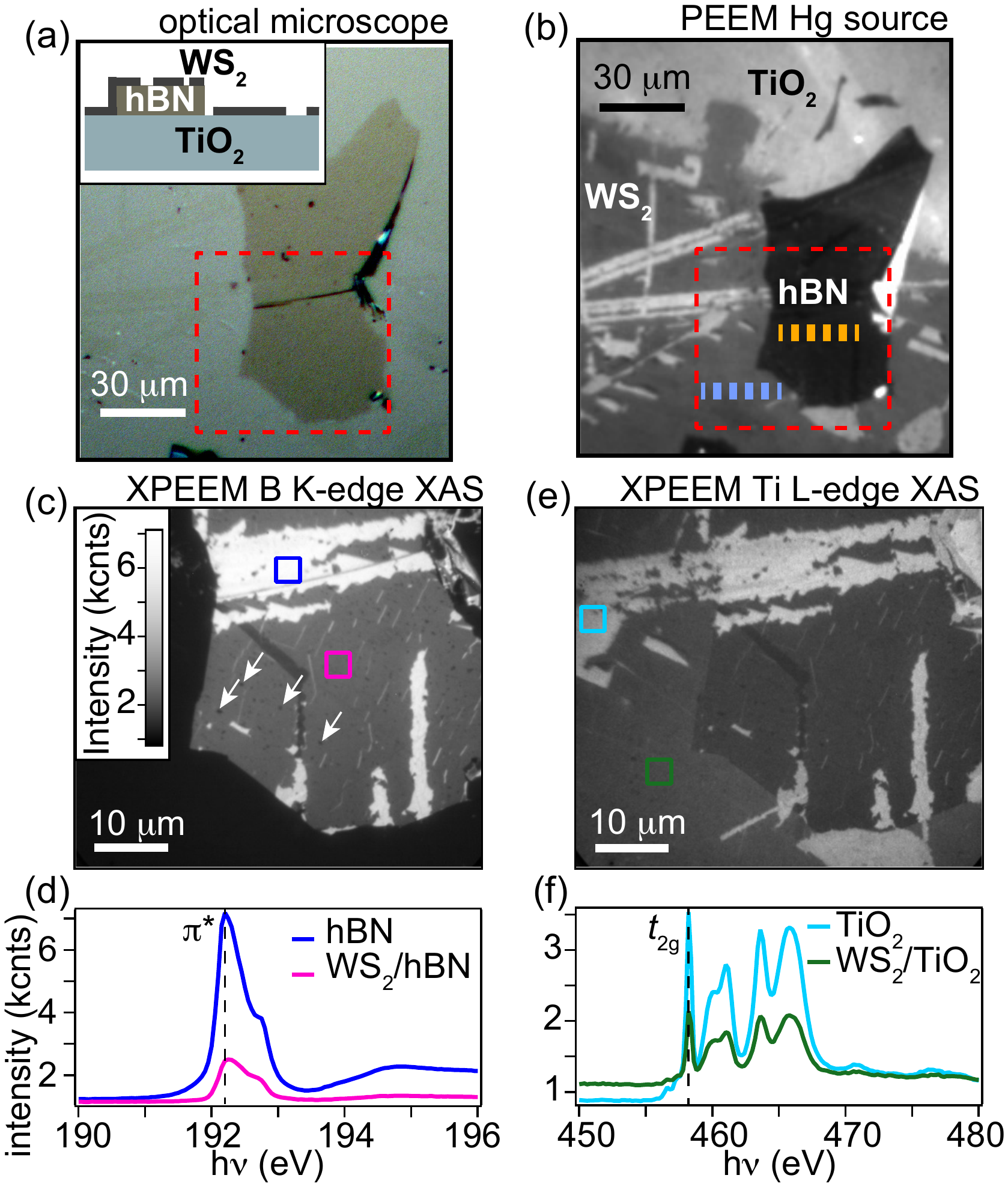}
\caption{X-ray absorption imaging: (a) Optical microscope image of a TiO$_2$ supported sample with the stacking  illustrated in the insert. (b) Hg PEEM image of the same region as shown in (a). (c) Secondary electron contrast from the $\pi^{\ast}$-resonance of the boron K-edge. White arrows point to examples of trapped bubbles in the van der Waals interface. (d) Area-selective XAS spectra of the boron K-edge collected from bare hBN (blue box in (c)) and WS$_2$ covered hBN (magenta box in (c)). (e)-(f) Corresponding (e) image and (f) XAS spectra for the $t_{2g}$ resonance of the Ti L-edge of bare TiO$_2$ (cyan box in (e)) and WS$_2$ on TiO$_2$ (green box in (e)). The color scale bar in (c) also applies to (e).}
\label{fig:2}
\end{center}
\end{figure}

The same piece of transferred SL WS$_2$ covers the SiO$_2$ substrate in the part marked by a dashed red box in Fig. \ref{fig:1}(a). We can therefore compare the contrast levels on both hBN and SiO$_2$ as shown in Figs. \ref{fig:1}(d)-(f). On WS$_2$/hBN in panel (d) the intensity exhibits only minor fluctuations with respect to the average, as demonstrated by the line profile in Fig. \ref{fig:1}(f). Much stronger contrasts are observed on WS$_2$/SiO$_2$ in panel (e), which are quantified in panel (f) as intensity fluctuations within a scale of 2 $\mu$m and a slow intensity increase over the full 30~$\mu$m range of the profile. These features are indicative of both long range and short range potential energy variations on the SiO$_2$, which are likely caused by remaining charge impurities that inevitably form in such WS$_2$/SiO$_2$ interfaces \cite{Ghatak:2011}. Removing such strong potential energy fluctuations is essential for electronic structure measurements as this greatly reduces energy broadening of the measured bands. This may be achieved using the conductive TiO$_2$ interface seen in the optical microscope image in Fig. \ref{fig:2}(a) and the Hg PEEM image in Fig. \ref{fig:2}(b) obtained after annealing to 380 $^{\circ}$C. Parts of a transferred WS$_2$ triangle straddle both the TiO$_2$ and the hBN flake. Representative line profiles from these two regions are compared with the SiO$_2$ sample in Fig. \ref{fig:1}(f) and exhibit much less fluctuations as expected for the conductive and thus more strongly screening TiO$_2$ interface \cite{Ulstrup:2016}.

X-ray PEEM (XPEEM) is applied for X-ray absorption spectroscopy (XAS) and imaging of the absorption peaks of the boron K-edge and titanium L-edge in Figs. \ref{fig:2}(c)-(f). The image in Fig. \ref{fig:2}(c) was obtained using secondary electron contrast of the boron $\pi^{\ast}$ resonance such that bare hBN areas exhibit a high intensity \cite{Koch:2018}. This reveals cracks and tears in the transferred WS$_2$ as well as dark sub-micron spots (see white arrows for a few examples in panel (c)) which are trapped bubbles that form in transferred van der Waals heterostructures \cite{Khestanova:2016}. The spatially resolved XAS spectra in Fig. \ref{fig:2}(d) are obtained by integrating the intensity within the blue and magenta boxes on bare and WS$_2$ covered hBN shown in panel (c). The expected $\pi^{\ast}$ resonance is observed in addition to a shoulder which appears after SL WS$_2$ transfer \cite{Koch:2018}. Using secondary electron contrast from the t$_{2g}$ resonance on the TiO$_2$ L-edge we are able to distinguish bare and WS$_2$ covered TiO$_2$ in Fig. \ref{fig:2}(e). The area-selective XAS spectra over the entire edge shown in Fig. \ref{fig:2}(f) resemble typical pristine TiO$_2$ spectra, indicating the cleanliness of the interface \cite{Ulstrup:2016}.

\begin{figure}
\begin{center}
\includegraphics[width=0.49\textwidth]{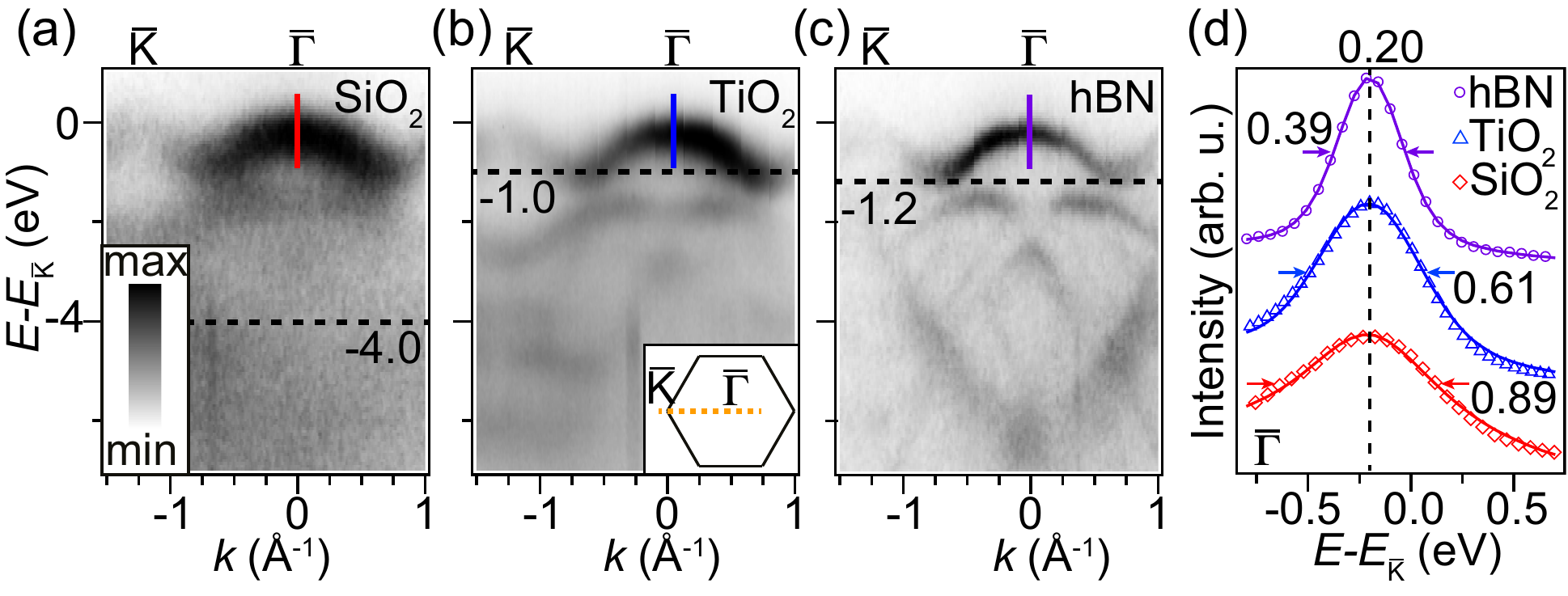}
\caption{Electronic structures measured by $k$-resolved PEEM: (a)-(c) Photoemission spectra of SL WS$_2$ on (a)  SiO$_2$, (b) TiO$_2$ and (c) hBN measured along the \kbar-\gbar~direction marked by an orange dashed line on the BZ in the insert of (b). The dashed horizontal lines in (a)-(c) provide the VB offsets for the substrates (error bars are $\pm 0.1$~eV). (d) EDCs (markers) with fits (curves) extracted at \gbar~as indicated by the correspondingly colored vertical bars in (a)-(c). The fitted peak position is marked with a vertical dashed line, and FWHM values are stated with arrows. All energy values are in units of eV.}
\label{fig:3}
\end{center}
\end{figure}

Having established the characteristic real space electronic contrasts, we collect distinct microARPES spectra with $k$-resolved PEEM from clean areas of the three vertical interfaces WS$_2$/SiO$_2$ (Fig. \ref{fig:3}(a)), WS$_2$/TiO$_2$ (Fig. \ref{fig:3}(b)) and WS$_2$/hBN (Fig. \ref{fig:3}(c)). The WS$_2$/hBN dispersion in Fig. \ref{fig:3}(c) is measured on the SiO$_2$ supported sample, but we get similar spectra from WS$_2$/hBN on TiO$_2$ \cite{Katoch:2018}. The data were obtained along the \kbar-\gbar~high symmetry direction of the SL WS$_2$ Brillouin zone (BZ), permitting us to identify the global valence band maximum (VBM) at \kbar~and the local maximum at \gbar~as expected for SL WS$_2$ \cite{Klein2001}. Note that the energy scale is referenced to the energy of the VBM at \kbar. Energy distribution curve (EDC) fits to Voigt line shapes on a linear background at \gbar~provide an offset of 0.20(4)~eV from the peak position to the VBM at \kbar~for all interfaces as seen in Fig. \ref{fig:3}(d). The full width at half maximum (FWHM) values for the fitted Voigt peaks demonstrate sharpest SL WS$_2$ bands on hBN with a FWHM value of 0.39(1)~eV (see arrows in Fig. \ref{fig:3}(d)). Extensive broadening is observed across the oxides with the FWHM value more than doubled on SiO$_2$.

\begin{figure}
\begin{center}
\includegraphics[width=0.49\textwidth]{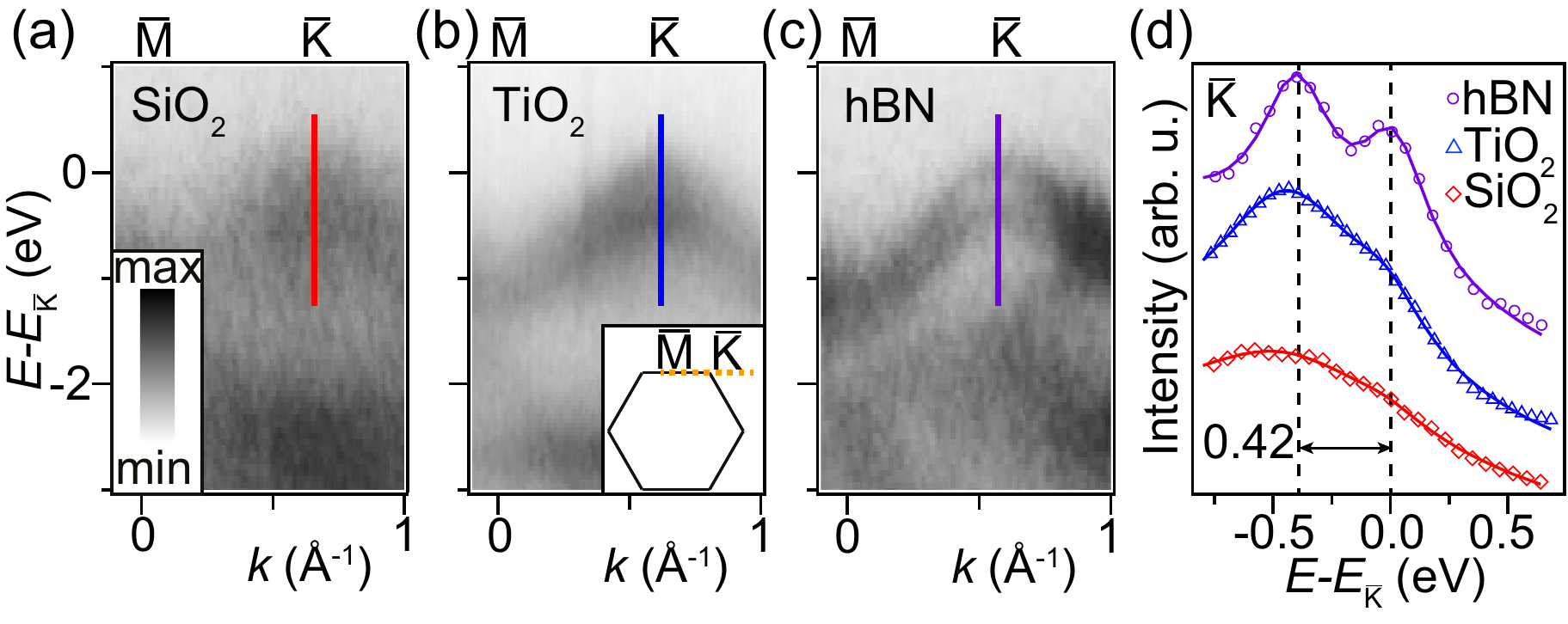}
\caption{Substrate influence on VBM at~\kbar~of SL WS$_2$: (a)-(c) Photoemission spectra on (a)  SiO$_2$, (b) TiO$_2$ and (c) hBN for the \mbar-\kbar~cut marked by an orange dashed line in (b). (d) EDCs (markers) with fits (curves) extracted at \kbar~as shown by correspondingly colored vertical bars in (a)-(c). The vertical dashed lines and double-headed arrow mark the given energy separation between the two peaks in units of eV.}
\label{fig:4}
\end{center}
\end{figure}

Measurements along \mbar-\kbar~further reveal the spin-orbit split VBs at \kbar~as seen in Figs. \ref{fig:4}(a)-(c). EDC fits lead to a value for the spin-orbit splitting of 0.42(6)~eV for WS$_2$/hBN as demonstrated in Fig. \ref{fig:4}(d), which is in agreement with other studies \cite{Ulstrup:2016,Katoch:2018}. The linewidth broadening masks the spin-orbit splitting to such an extent that the EDC fits for SiO$_2$ and TiO$_2$ in Fig. \ref{fig:3}(g) had to be performed with the peak separations constrained to the values obtained on hBN. The broad VB states of WS$_2$/SiO$_2$ are consistent with similar measurements on MoS$_2$/SiO$_2$  \cite{Jin2013,Jin:2015,Yuan:2016}, which may be explained by charge impurities rigidly shifting and broadening the bands as hinted by the work function contrast in Fig. \ref{fig:1}(e). Such effects are also present in TiO$_2$, although less dramatic \cite{Ulstrup:2016}. The surface roughness in the oxides is expected to be substantially higher than in hBN~\cite{lui2009}, which causes additional momentum broadening. 

\begin{figure}
\begin{center}
\includegraphics[width=0.49\textwidth]{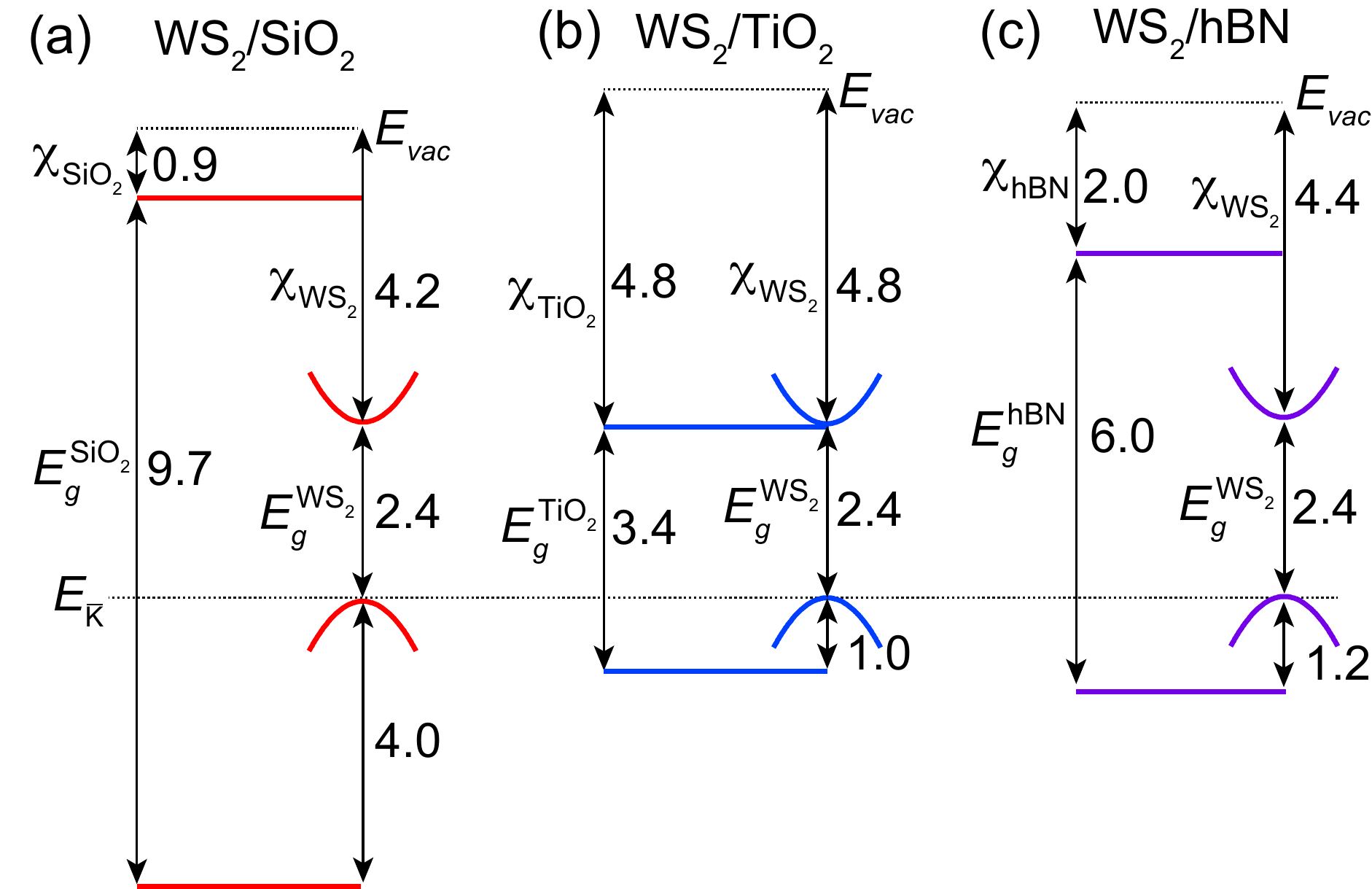}
\caption{SL WS$_2$ band alignments with (a) SiO$_2$, (b) TiO$_2$ and (c) hBN. The SL WS$_2$ VBM at \kbar~is used as a general reference for the three systems (see horizontal dotted line). All values are given in units of eV. The error bars on the VB offsets and $\chi_{\mathrm{WS_2}}$ are $\pm 0.1$ eV. Literature values for $E_g$ and $\chi$ are summarized in Table \ref{tab:1}.}
\label{fig:5}
\end{center}
\end{figure}

We determine the VBM offsets for the substrates (marked by dashed horizontal lines in Figs. \ref{fig:3}(a)-(c)) as described in the Supplementary Material and apply the electron affinity rule as the simplest method of constructing the band alignment diagrams of our mixed 2D-3D heterojunctions with respect to the vacuum level $E_{vac}$ in Fig. \ref{fig:5} \cite{Klein2012}. In all cases we assume the measured quasiparticle band gap of SL WS$_2$ on SiO$_2$ given by $E^{\mathrm{WS_2}}_g = 2.4$~eV~\cite{Chernikov:2015b}. Substrate values for $E_g$ and $\chi$ are given in Fig. \ref{fig:5} and Table \ref{tab:1}. On both SiO$_2$ and hBN a straddling band gap configuration appears due to the wide gaps of the substrates (see panels (a) and (c)). On TiO$_2$ the conduction band offsets are very close and may form a staggered band gap (panel (b)), which could lead to substantial electron (hole) transfer to TiO$_2$ (SL WS$_2$). This may explain our previous observation of less electron doping of SL WS$_2$ on TiO$_2$ compared to other oxides \cite{Ulstrup:2016}. 

\begin{table}
  \centering
  \begin{tabular*}{0.49\textwidth}{@{\extracolsep{\fill} } l c c c c }
  \hline		
    \hline	
   Substrate  & $\epsilon$ & $E_g$ (eV) & $\chi$ (eV) & $E-E_{\mathrm{\bar{K}}}^{\mathrm{WS}_2}$ (eV)\\
  \hline	
  SiO$_2$		&		3.9		&		9.7 \cite{Kresse:2012}		&		0.9 \cite{Williams:1965}	&		4.0 \\
  TiO$_2$	&		113		&		3.4 \cite{Landmann:2012}	&		4.8 \cite{Scanlon:2013}	&		1.0 \\
  hBN	&		4.0		&		6.0 \cite{Arnaud:2006}	&		2.0 \cite{Choi:2013}	&		1.2  \\
  \hline  
  \hline
\end{tabular*}
  \caption{Summary of dielectric constants $\epsilon$, quasiparticle band gaps $E_g$, electron affinities $\chi$ and band offsets with respect to \kbar~of SL WS$_2$ for SiO$_2$, TiO$_2$ and hBN.}
   \label{tab:1}
\end{table}

This simple construction suggests that $\chi_{\mathrm{WS_2}}$ is substrate dependent and generally larger than a recently determined theoretical value of 3.75~eV~\cite{Guo:2016}. Caution should be exercised when considering the values here because of the variation in literature values of $\chi$ for the substrates. This issue is most pronounced in the case of $\chi_{hBN}$ where we used an often cited value of 2.0 eV \cite{Choi:2013} in Fig. \ref{fig:5}(c). However, a value of 1.1~eV can also be found \cite{Fiori:2012} and even a negative $\chi_{hBN}$ has been suggested \cite{Powers:1995}. Note also that on TiO$_2$ the Nb doping as well as annealing- and beam-induced oxygen vacancies may modify the band offsets from their intrinsic values \cite{Onda2004}, which could lead to an overestimation of $\chi_{\mathrm{WS_2}}$. Additionally, the simple electron affinity rule may break down due to a substrate dependent quasiparticle band gap of SL WS$_2$ or possibly due to unusually strong interfacial dipoles that vary between substrates \cite{Schlaf:1999}. 

In conclusion, we have fabricated SL WS$_2$/hBN heterostructures supported on SiO$_2$ and TiO$_2$ substrates implementing device architectures in photoemission spectromicroscopy experiments that we believe will be compatible with charge transport measurements in gated conditions and with current passing through the materials \cite{Kaminski:2016}. The electronic transport properties of these mixed 2D-3D junctions will be defined by the vertical band alignments which we here inferred using an electron affinity rule incorporating the measured VB offsets.

\section*{supplementary material}
See online Supplementary Material for further details on PEEM measurements, for the sample fabrication procedure and for the determination of the SiO$_2$, TiO$_2$ and hBN VB offsets.

\section*{acknowledgement}
S. U. acknowledges financial support from VILLUM FONDEN (Grant. No. 15375). R. J. K. is supported by a fellowship within the Postdoc-Program of the German Academic Exchange Service (DAAD). D. S. acknowledges financial support from the Netherlands Organisation for Scientific Research under the Rubicon Program (Grant 680-50-1305). The Advanced Light Source is supported by the Director, Office of Science, Office of Basic Energy Sciences, of the U.S. Department of Energy under Contract No. DE-AC02-05CH11231. This work was supported by IBS-R009-D1. The work at NRL was supported by core programs and the Nanoscience Institute.


\begin{thebibliography}{39}%
\makeatletter
\providecommand \@ifxundefined [1]{%
 \@ifx{#1\undefined}
}%
\providecommand \@ifnum [1]{%
 \ifnum #1\expandafter \@firstoftwo
 \else \expandafter \@secondoftwo
 \fi
}%
\providecommand \@ifx [1]{%
 \ifx #1\expandafter \@firstoftwo
 \else \expandafter \@secondoftwo
 \fi
}%
\providecommand \natexlab [1]{#1}%
\providecommand \enquote  [1]{``#1''}%
\providecommand \bibnamefont  [1]{#1}%
\providecommand \bibfnamefont [1]{#1}%
\providecommand \citenamefont [1]{#1}%
\providecommand \href@noop [0]{\@secondoftwo}%
\providecommand \href [0]{\begingroup \@sanitize@url \@href}%
\providecommand \@href[1]{\@@startlink{#1}\@@href}%
\providecommand \@@href[1]{\endgroup#1\@@endlink}%
\providecommand \@sanitize@url [0]{\catcode `\\12\catcode `\$12\catcode
  `\&12\catcode `\#12\catcode `\^12\catcode `\_12\catcode `\%12\relax}%
\providecommand \@@startlink[1]{}%
\providecommand \@@endlink[0]{}%
\providecommand \url  [0]{\begingroup\@sanitize@url \@url }%
\providecommand \@url [1]{\endgroup\@href {#1}{\urlprefix }}%
\providecommand \urlprefix  [0]{URL }%
\providecommand \Eprint [0]{\href }%
\providecommand \doibase [0]{http://dx.doi.org/}%
\providecommand \selectlanguage [0]{\@gobble}%
\providecommand \bibinfo  [0]{\@secondoftwo}%
\providecommand \bibfield  [0]{\@secondoftwo}%
\providecommand \translation [1]{[#1]}%
\providecommand \BibitemOpen [0]{}%
\providecommand \bibitemStop [0]{}%
\providecommand \bibitemNoStop [0]{.\EOS\space}%
\providecommand \EOS [0]{\spacefactor3000\relax}%
\providecommand \BibitemShut  [1]{\csname bibitem#1\endcsname}%
\let\auto@bib@innerbib\@empty
%</preamble>
\bibitem [{\citenamefont {Radisavljevic}\ \emph {et~al.}(2011)\citenamefont
  {Radisavljevic}, \citenamefont {Radenovic}, \citenamefont {Brivio},
  \citenamefont {Giacometti},\ and\ \citenamefont {Kis}}]{radisavljevic2011}%
  \BibitemOpen
  \bibfield  {author} {\bibinfo {author} {\bibfnamefont {B.}~\bibnamefont
  {Radisavljevic}}, \bibinfo {author} {\bibfnamefont {A.}~\bibnamefont
  {Radenovic}}, \bibinfo {author} {\bibfnamefont {J.}~\bibnamefont {Brivio}},
  \bibinfo {author} {\bibfnamefont {V.}~\bibnamefont {Giacometti}}, \ and\
  \bibinfo {author} {\bibfnamefont {A.}~\bibnamefont {Kis}},\ }\href {\doibase
  10.1038/NNANO.2010.279} {\bibfield  {journal} {\bibinfo  {journal} {Nature
  nanotechnology}\ }\textbf {\bibinfo {volume} {6}},\ \bibinfo {pages} {147}
  (\bibinfo {year} {2011})}\BibitemShut {NoStop}%
\bibitem [{\citenamefont {Wang}\ \emph {et~al.}(2012)\citenamefont {Wang},
  \citenamefont {Kourosh}, \citenamefont {Kis}, \citenamefont {Coleman},\ and\
  \citenamefont {Strano}}]{wang2012}%
  \BibitemOpen
  \bibfield  {author} {\bibinfo {author} {\bibfnamefont {Q.}~\bibnamefont
  {Wang}}, \bibinfo {author} {\bibfnamefont {K.}~\bibnamefont {Kourosh}},
  \bibinfo {author} {\bibfnamefont {A.}~\bibnamefont {Kis}}, \bibinfo {author}
  {\bibfnamefont {J.}~\bibnamefont {Coleman}}, \ and\ \bibinfo {author}
  {\bibfnamefont {M.}~\bibnamefont {Strano}},\ }\href {\doibase
  10.1038/nnano.2012.193} {\bibfield  {journal} {\bibinfo  {journal} {Nature
  Nanotechnology}\ }\textbf {\bibinfo {volume} {7}},\ \bibinfo {pages} {699}
  (\bibinfo {year} {2012})}\BibitemShut {NoStop}%
\bibitem [{\citenamefont {Jariwala}\ \emph {et~al.}(2013)\citenamefont
  {Jariwala}, \citenamefont {Sangwan}, \citenamefont {Late}, \citenamefont
  {Johns}, \citenamefont {Dravid}, \citenamefont {Marks}, \citenamefont
  {Lauhon},\ and\ \citenamefont {Hersam}}]{jariwala2013}%
  \BibitemOpen
  \bibfield  {author} {\bibinfo {author} {\bibfnamefont {D.}~\bibnamefont
  {Jariwala}}, \bibinfo {author} {\bibfnamefont {V.~K.}\ \bibnamefont
  {Sangwan}}, \bibinfo {author} {\bibfnamefont {D.~J.}\ \bibnamefont {Late}},
  \bibinfo {author} {\bibfnamefont {J.~E.}\ \bibnamefont {Johns}}, \bibinfo
  {author} {\bibfnamefont {V.~P.}\ \bibnamefont {Dravid}}, \bibinfo {author}
  {\bibfnamefont {T.~J.}\ \bibnamefont {Marks}}, \bibinfo {author}
  {\bibfnamefont {L.~J.}\ \bibnamefont {Lauhon}}, \ and\ \bibinfo {author}
  {\bibfnamefont {M.~C.}\ \bibnamefont {Hersam}},\ }\href@noop {} {\bibfield
  {journal} {\bibinfo  {journal} {Appl. Phys. Lett.}\ }\textbf {\bibinfo
  {volume} {102}},\ \bibinfo {pages} {173107} (\bibinfo {year}
  {2013})}\BibitemShut {NoStop}%
\bibitem [{\citenamefont {Jariwala}\ \emph {et~al.}(2014)\citenamefont
  {Jariwala}, \citenamefont {Sangwan}, \citenamefont {Lauhon}, \citenamefont
  {Marks},\ and\ \citenamefont {Hersam}}]{jariwala2014}%
  \BibitemOpen
  \bibfield  {author} {\bibinfo {author} {\bibfnamefont {D.}~\bibnamefont
  {Jariwala}}, \bibinfo {author} {\bibfnamefont {V.~K.}\ \bibnamefont
  {Sangwan}}, \bibinfo {author} {\bibfnamefont {L.~J.}\ \bibnamefont {Lauhon}},
  \bibinfo {author} {\bibfnamefont {T.~J.}\ \bibnamefont {Marks}}, \ and\
  \bibinfo {author} {\bibfnamefont {M.~C.}\ \bibnamefont {Hersam}},\
  }\href@noop {} {\bibfield  {journal} {\bibinfo  {journal} {ACS Nano}\
  }\textbf {\bibinfo {volume} {8}},\ \bibinfo {pages} {1102} (\bibinfo {year}
  {2014})}\BibitemShut {NoStop}%
\bibitem [{\citenamefont {Gong}\ \emph {et~al.}(2014)\citenamefont {Gong},
  \citenamefont {Lin}, \citenamefont {Wang}, \citenamefont {Shi}, \citenamefont
  {Lei}, \citenamefont {Lin}, \citenamefont {Zou}, \citenamefont {Ye},
  \citenamefont {Vajtai}, \citenamefont {Yakobson}, \citenamefont {Terrones},
  \citenamefont {Terrones}, \citenamefont {Tay}, \citenamefont {Lou},
  \citenamefont {Pantelides}, \citenamefont {Liu}, \citenamefont {Zhou},\ and\
  \citenamefont {Ajayan}}]{Gong:2014}%
  \BibitemOpen
  \bibfield  {author} {\bibinfo {author} {\bibfnamefont {Y.}~\bibnamefont
  {Gong}}, \bibinfo {author} {\bibfnamefont {J.}~\bibnamefont {Lin}}, \bibinfo
  {author} {\bibfnamefont {X.}~\bibnamefont {Wang}}, \bibinfo {author}
  {\bibfnamefont {G.}~\bibnamefont {Shi}}, \bibinfo {author} {\bibfnamefont
  {S.}~\bibnamefont {Lei}}, \bibinfo {author} {\bibfnamefont {Z.}~\bibnamefont
  {Lin}}, \bibinfo {author} {\bibfnamefont {X.}~\bibnamefont {Zou}}, \bibinfo
  {author} {\bibfnamefont {G.}~\bibnamefont {Ye}}, \bibinfo {author}
  {\bibfnamefont {R.}~\bibnamefont {Vajtai}}, \bibinfo {author} {\bibfnamefont
  {B.~I.}\ \bibnamefont {Yakobson}}, \bibinfo {author} {\bibfnamefont
  {H.}~\bibnamefont {Terrones}}, \bibinfo {author} {\bibfnamefont
  {M.}~\bibnamefont {Terrones}}, \bibinfo {author} {\bibfnamefont {B.~K.}\
  \bibnamefont {Tay}}, \bibinfo {author} {\bibfnamefont {J.}~\bibnamefont
  {Lou}}, \bibinfo {author} {\bibfnamefont {S.~T.}\ \bibnamefont {Pantelides}},
  \bibinfo {author} {\bibfnamefont {Z.}~\bibnamefont {Liu}}, \bibinfo {author}
  {\bibfnamefont {W.}~\bibnamefont {Zhou}}, \ and\ \bibinfo {author}
  {\bibfnamefont {P.~M.}\ \bibnamefont {Ajayan}},\ }\href
  {https://doi.org/10.1038/nmat4091} {\bibfield  {journal} {\bibinfo  {journal}
  {Nature Materials}\ }\textbf {\bibinfo {volume} {13}},\ \bibinfo {pages}
  {1135} (\bibinfo {year} {2014})}\BibitemShut {NoStop}%
\bibitem [{\citenamefont {Allain}\ \emph {et~al.}(2015)\citenamefont {Allain},
  \citenamefont {Kang}, \citenamefont {Banerjee},\ and\ \citenamefont
  {Kis}}]{Allain2015}%
  \BibitemOpen
  \bibfield  {author} {\bibinfo {author} {\bibfnamefont {A.}~\bibnamefont
  {Allain}}, \bibinfo {author} {\bibfnamefont {J.}~\bibnamefont {Kang}},
  \bibinfo {author} {\bibfnamefont {K.}~\bibnamefont {Banerjee}}, \ and\
  \bibinfo {author} {\bibfnamefont {A.}~\bibnamefont {Kis}},\ }\href
  {https://doi.org/10.1038/nmat4452} {\bibfield  {journal} {\bibinfo  {journal}
  {Nature Materials}\ }\textbf {\bibinfo {volume} {14}},\ \bibinfo {pages}
  {1195} (\bibinfo {year} {2015})}\BibitemShut {NoStop}%
\bibitem [{\citenamefont {Schlaf}\ \emph {et~al.}(1999)\citenamefont {Schlaf},
  \citenamefont {Lang}, \citenamefont {Pettenkofer},\ and\ \citenamefont
  {Jaegermann}}]{Schlaf:1999}%
  \BibitemOpen
  \bibfield  {author} {\bibinfo {author} {\bibfnamefont {R.}~\bibnamefont
  {Schlaf}}, \bibinfo {author} {\bibfnamefont {O.}~\bibnamefont {Lang}},
  \bibinfo {author} {\bibfnamefont {C.}~\bibnamefont {Pettenkofer}}, \ and\
  \bibinfo {author} {\bibfnamefont {W.}~\bibnamefont {Jaegermann}},\ }\href
  {\doibase 10.1063/1.369590} {\bibfield  {journal} {\bibinfo  {journal}
  {Journal of Applied Physics}\ }\textbf {\bibinfo {volume} {85}},\ \bibinfo
  {pages} {2732} (\bibinfo {year} {1999})}\BibitemShut {NoStop}%
\bibitem [{\citenamefont {McDonnell}\ \emph {et~al.}(2014)\citenamefont
  {McDonnell}, \citenamefont {Azcatl}, \citenamefont {Addou}, \citenamefont
  {Gong}, \citenamefont {Battaglia}, \citenamefont {Chuang}, \citenamefont
  {Cho}, \citenamefont {Javey},\ and\ \citenamefont
  {Wallace}}]{McDonnell:2014}%
  \BibitemOpen
  \bibfield  {author} {\bibinfo {author} {\bibfnamefont {S.}~\bibnamefont
  {McDonnell}}, \bibinfo {author} {\bibfnamefont {A.}~\bibnamefont {Azcatl}},
  \bibinfo {author} {\bibfnamefont {R.}~\bibnamefont {Addou}}, \bibinfo
  {author} {\bibfnamefont {C.}~\bibnamefont {Gong}}, \bibinfo {author}
  {\bibfnamefont {C.}~\bibnamefont {Battaglia}}, \bibinfo {author}
  {\bibfnamefont {S.}~\bibnamefont {Chuang}}, \bibinfo {author} {\bibfnamefont
  {K.}~\bibnamefont {Cho}}, \bibinfo {author} {\bibfnamefont {A.}~\bibnamefont
  {Javey}}, \ and\ \bibinfo {author} {\bibfnamefont {R.~M.}\ \bibnamefont
  {Wallace}},\ }\href {\doibase 10.1021/nn501728w} {\bibfield  {journal}
  {\bibinfo  {journal} {ACS Nano}\ }\textbf {\bibinfo {volume} {8}},\ \bibinfo
  {pages} {6265} (\bibinfo {year} {2014})}\BibitemShut {NoStop}%
\bibitem [{\citenamefont {Ulstrup}\ \emph {et~al.}(2016)\citenamefont
  {Ulstrup}, \citenamefont {Katoch}, \citenamefont {Koch}, \citenamefont
  {Schwarz}, \citenamefont {Singh}, \citenamefont {Mc{C}reary}, \citenamefont
  {Yoo}, \citenamefont {Xu}, \citenamefont {Jonker}, \citenamefont {Kawakami},
  \citenamefont {Bostwick}, \citenamefont {Rotenberg},\ and\ \citenamefont
  {Jozwiak}}]{Ulstrup:2016}%
  \BibitemOpen
  \bibfield  {author} {\bibinfo {author} {\bibfnamefont {S.}~\bibnamefont
  {Ulstrup}}, \bibinfo {author} {\bibfnamefont {J.}~\bibnamefont {Katoch}},
  \bibinfo {author} {\bibfnamefont {R.}~\bibnamefont {Koch}}, \bibinfo {author}
  {\bibfnamefont {D.}~\bibnamefont {Schwarz}}, \bibinfo {author} {\bibfnamefont
  {S.}~\bibnamefont {Singh}}, \bibinfo {author} {\bibfnamefont
  {K.}~\bibnamefont {Mc{C}reary}}, \bibinfo {author} {\bibfnamefont
  {H.}~\bibnamefont {Yoo}}, \bibinfo {author} {\bibfnamefont {J.}~\bibnamefont
  {Xu}}, \bibinfo {author} {\bibfnamefont {B.}~\bibnamefont {Jonker}}, \bibinfo
  {author} {\bibfnamefont {R.}~\bibnamefont {Kawakami}}, \bibinfo {author}
  {\bibfnamefont {A.}~\bibnamefont {Bostwick}}, \bibinfo {author}
  {\bibfnamefont {E.}~\bibnamefont {Rotenberg}}, \ and\ \bibinfo {author}
  {\bibfnamefont {C.}~\bibnamefont {Jozwiak}},\ }\href {\doibase
  10.1021/acsnano.6b04914} {\bibfield  {journal} {\bibinfo  {journal} {ACS
  Nano}\ }\textbf {\bibinfo {volume} {10}},\ \bibinfo {pages} {10058} (\bibinfo
  {year} {2016})}\BibitemShut {NoStop}%
\bibitem [{\citenamefont {Jariwala}\ \emph {et~al.}(2016)\citenamefont
  {Jariwala}, \citenamefont {Marks},\ and\ \citenamefont
  {Hersam}}]{Jariwala:2016}%
  \BibitemOpen
  \bibfield  {author} {\bibinfo {author} {\bibfnamefont {D.}~\bibnamefont
  {Jariwala}}, \bibinfo {author} {\bibfnamefont {T.~J.}\ \bibnamefont {Marks}},
  \ and\ \bibinfo {author} {\bibfnamefont {M.~C.}\ \bibnamefont {Hersam}},\
  }\href {https://doi.org/10.1038/nmat4703} {\bibfield  {journal} {\bibinfo
  {journal} {Nature Materials}\ }\textbf {\bibinfo {volume} {16}},\ \bibinfo
  {pages} {170} (\bibinfo {year} {2016})}\BibitemShut {NoStop}%
\bibitem [{\citenamefont {Guo}\ and\ \citenamefont
  {Robertson}(2016)}]{Guo:2016}%
  \BibitemOpen
  \bibfield  {author} {\bibinfo {author} {\bibfnamefont {Y.}~\bibnamefont
  {Guo}}\ and\ \bibinfo {author} {\bibfnamefont {J.}~\bibnamefont
  {Robertson}},\ }\href {\doibase 10.1063/1.4953169} {\bibfield  {journal}
  {\bibinfo  {journal} {Applied Physics Letters}\ }\textbf {\bibinfo {volume}
  {108}},\ \bibinfo {pages} {233104} (\bibinfo {year} {2016})}\BibitemShut
  {NoStop}%
\bibitem [{\citenamefont {Klein}(2012)}]{Klein2012}%
  \BibitemOpen
  \bibfield  {author} {\bibinfo {author} {\bibfnamefont {A.}~\bibnamefont
  {Klein}},\ }\href {\doibase https://doi.org/10.1016/j.tsf.2011.10.055}
  {\bibfield  {journal} {\bibinfo  {journal} {Thin Solid Films}\ }\textbf
  {\bibinfo {volume} {520}},\ \bibinfo {pages} {3721 } (\bibinfo {year}
  {2012})},\ \bibinfo {note} {7th International Symposium on Transparent Oxide
  Thin Films for Electronics and Optics (TOEO-7)}\BibitemShut {NoStop}%
\bibitem [{\citenamefont {Fujikawa}\ \emph {et~al.}(2009)\citenamefont
  {Fujikawa}, \citenamefont {Sakurai},\ and\ \citenamefont
  {Tromp}}]{Fujikawa2009}%
  \BibitemOpen
  \bibfield  {author} {\bibinfo {author} {\bibfnamefont {Y.}~\bibnamefont
  {Fujikawa}}, \bibinfo {author} {\bibfnamefont {T.}~\bibnamefont {Sakurai}}, \
  and\ \bibinfo {author} {\bibfnamefont {R.~M.}\ \bibnamefont {Tromp}},\
  }\href@noop {} {\bibfield  {journal} {\bibinfo  {journal} {Phys. Rev. B}\
  }\textbf {\bibinfo {volume} {79}},\ \bibinfo {pages} {121401(R)} (\bibinfo
  {year} {2009})}\BibitemShut {NoStop}%
\bibitem [{\citenamefont {Koch}\ \emph {et~al.}(2018)\citenamefont {Koch},
  \citenamefont {Katoch}, \citenamefont {Moser}, \citenamefont {Schwarz},
  \citenamefont {Kawakami}, \citenamefont {Bostwick}, \citenamefont
  {Rotenberg}, \citenamefont {Jozwiak},\ and\ \citenamefont
  {Ulstrup}}]{Koch:2018}%
  \BibitemOpen
  \bibfield  {author} {\bibinfo {author} {\bibfnamefont {R.~J.}\ \bibnamefont
  {Koch}}, \bibinfo {author} {\bibfnamefont {J.}~\bibnamefont {Katoch}},
  \bibinfo {author} {\bibfnamefont {S.}~\bibnamefont {Moser}}, \bibinfo
  {author} {\bibfnamefont {D.}~\bibnamefont {Schwarz}}, \bibinfo {author}
  {\bibfnamefont {R.~K.}\ \bibnamefont {Kawakami}}, \bibinfo {author}
  {\bibfnamefont {A.}~\bibnamefont {Bostwick}}, \bibinfo {author}
  {\bibfnamefont {E.}~\bibnamefont {Rotenberg}}, \bibinfo {author}
  {\bibfnamefont {C.}~\bibnamefont {Jozwiak}}, \ and\ \bibinfo {author}
  {\bibfnamefont {S.}~\bibnamefont {Ulstrup}},\ }\href {\doibase
  10.1103/PhysRevMaterials.2.074006} {\bibfield  {journal} {\bibinfo  {journal}
  {Phys. Rev. Materials}\ }\textbf {\bibinfo {volume} {2}},\ \bibinfo {pages}
  {074006} (\bibinfo {year} {2018})}\BibitemShut {NoStop}%
\bibitem [{\citenamefont {Jin}\ \emph {et~al.}(2013)\citenamefont {Jin},
  \citenamefont {Yeh}, \citenamefont {Zaki}, \citenamefont {Zhang},
  \citenamefont {Sadowski}, \citenamefont {Abdullah}, \citenamefont {Zande},
  \citenamefont {Chenet}, \citenamefont {Dadap}, \citenamefont {Herman},
  \citenamefont {Sutter}, \citenamefont {Hone},\ and\ \citenamefont
  {Osgood}}]{Jin2013}%
  \BibitemOpen
  \bibfield  {author} {\bibinfo {author} {\bibfnamefont {W.}~\bibnamefont
  {Jin}}, \bibinfo {author} {\bibfnamefont {P.}~\bibnamefont {Yeh}}, \bibinfo
  {author} {\bibfnamefont {N.}~\bibnamefont {Zaki}}, \bibinfo {author}
  {\bibfnamefont {D.}~\bibnamefont {Zhang}}, \bibinfo {author} {\bibfnamefont
  {J.}~\bibnamefont {Sadowski}}, \bibinfo {author} {\bibfnamefont
  {A.}~\bibnamefont {Abdullah}}, \bibinfo {author} {\bibfnamefont
  {A.}~\bibnamefont {Zande}}, \bibinfo {author} {\bibfnamefont
  {D.}~\bibnamefont {Chenet}}, \bibinfo {author} {\bibfnamefont
  {J.}~\bibnamefont {Dadap}}, \bibinfo {author} {\bibfnamefont
  {I.}~\bibnamefont {Herman}}, \bibinfo {author} {\bibfnamefont
  {P.}~\bibnamefont {Sutter}}, \bibinfo {author} {\bibfnamefont
  {J.}~\bibnamefont {Hone}}, \ and\ \bibinfo {author} {\bibfnamefont
  {R.}~\bibnamefont {Osgood}},\ }\href@noop {} {\bibfield  {journal} {\bibinfo
  {journal} {Physical Review Letters}\ }\textbf {\bibinfo {volume} {111}},\
  \bibinfo {pages} {106801} (\bibinfo {year} {2013})}\BibitemShut {NoStop}%
\bibitem [{\citenamefont {Jin}\ \emph {et~al.}(2015)\citenamefont {Jin},
  \citenamefont {Yeh}, \citenamefont {Zaki}, \citenamefont {Zhang},
  \citenamefont {Liou}, \citenamefont {Sadowski}, \citenamefont {Barinov},
  \citenamefont {Yablonskikh}, \citenamefont {Dadap}, \citenamefont {Sutter},
  \citenamefont {Herman},\ and\ \citenamefont {Osgood}}]{Jin:2015}%
  \BibitemOpen
  \bibfield  {author} {\bibinfo {author} {\bibfnamefont {W.}~\bibnamefont
  {Jin}}, \bibinfo {author} {\bibfnamefont {P.-C.}\ \bibnamefont {Yeh}},
  \bibinfo {author} {\bibfnamefont {N.}~\bibnamefont {Zaki}}, \bibinfo {author}
  {\bibfnamefont {D.}~\bibnamefont {Zhang}}, \bibinfo {author} {\bibfnamefont
  {J.~T.}\ \bibnamefont {Liou}}, \bibinfo {author} {\bibfnamefont {J.~T.}\
  \bibnamefont {Sadowski}}, \bibinfo {author} {\bibfnamefont {A.}~\bibnamefont
  {Barinov}}, \bibinfo {author} {\bibfnamefont {M.}~\bibnamefont
  {Yablonskikh}}, \bibinfo {author} {\bibfnamefont {J.~I.}\ \bibnamefont
  {Dadap}}, \bibinfo {author} {\bibfnamefont {P.}~\bibnamefont {Sutter}},
  \bibinfo {author} {\bibfnamefont {I.~P.}\ \bibnamefont {Herman}}, \ and\
  \bibinfo {author} {\bibfnamefont {R.~M.}\ \bibnamefont {Osgood}},\ }\href
  {\doibase 10.1103/PhysRevB.91.121409} {\bibfield  {journal} {\bibinfo
  {journal} {Phys. Rev. B}\ }\textbf {\bibinfo {volume} {91}},\ \bibinfo
  {pages} {121409} (\bibinfo {year} {2015})}\BibitemShut {NoStop}%
\bibitem [{\citenamefont {Yeh}\ \emph {et~al.}(2015)\citenamefont {Yeh},
  \citenamefont {Jin}, \citenamefont {Zaki}, \citenamefont {Zhang},
  \citenamefont {Liou}, \citenamefont {Sadowski}, \citenamefont {Al-Mahboob},
  \citenamefont {Dadap}, \citenamefont {Herman}, \citenamefont {Sutter},\ and\
  \citenamefont {Osgood}}]{Yeh:2015}%
  \BibitemOpen
  \bibfield  {author} {\bibinfo {author} {\bibfnamefont {P.-C.}\ \bibnamefont
  {Yeh}}, \bibinfo {author} {\bibfnamefont {W.}~\bibnamefont {Jin}}, \bibinfo
  {author} {\bibfnamefont {N.}~\bibnamefont {Zaki}}, \bibinfo {author}
  {\bibfnamefont {D.}~\bibnamefont {Zhang}}, \bibinfo {author} {\bibfnamefont
  {J.~T.}\ \bibnamefont {Liou}}, \bibinfo {author} {\bibfnamefont {J.~T.}\
  \bibnamefont {Sadowski}}, \bibinfo {author} {\bibfnamefont {A.}~\bibnamefont
  {Al-Mahboob}}, \bibinfo {author} {\bibfnamefont {J.~I.}\ \bibnamefont
  {Dadap}}, \bibinfo {author} {\bibfnamefont {I.~P.}\ \bibnamefont {Herman}},
  \bibinfo {author} {\bibfnamefont {P.}~\bibnamefont {Sutter}}, \ and\ \bibinfo
  {author} {\bibfnamefont {R.~M.}\ \bibnamefont {Osgood}},\ }\href {\doibase
  10.1103/PhysRevB.91.041407} {\bibfield  {journal} {\bibinfo  {journal} {Phys.
  Rev. B}\ }\textbf {\bibinfo {volume} {91}},\ \bibinfo {pages} {041407}
  (\bibinfo {year} {2015})}\BibitemShut {NoStop}%
\bibitem [{\citenamefont {McCreary}\ \emph {et~al.}(2016)\citenamefont
  {McCreary}, \citenamefont {Hanbicki}, \citenamefont {Jernigan}, \citenamefont
  {Culbertson},\ and\ \citenamefont {Jonker}}]{McCreary:2016}%
  \BibitemOpen
  \bibfield  {author} {\bibinfo {author} {\bibfnamefont {K.~M.}\ \bibnamefont
  {McCreary}}, \bibinfo {author} {\bibfnamefont {A.~T.}\ \bibnamefont
  {Hanbicki}}, \bibinfo {author} {\bibfnamefont {G.~G.}\ \bibnamefont
  {Jernigan}}, \bibinfo {author} {\bibfnamefont {J.~C.}\ \bibnamefont
  {Culbertson}}, \ and\ \bibinfo {author} {\bibfnamefont {B.~T.}\ \bibnamefont
  {Jonker}},\ }\href {https://doi.org/10.1038/srep19159} {\bibfield  {journal}
  {\bibinfo  {journal} {Scientific Reports}\ }\textbf {\bibinfo {volume} {6}},\
  \bibinfo {pages} {19159} (\bibinfo {year} {2016})}\BibitemShut {NoStop}%
\bibitem [{\citenamefont {Katoch}\ \emph {et~al.}(2018)\citenamefont {Katoch},
  \citenamefont {Ulstrup}, \citenamefont {Koch}, \citenamefont {Moser},
  \citenamefont {Mc{C}reary}, \citenamefont {Singh}, \citenamefont {Xu},
  \citenamefont {Jonker}, \citenamefont {Kawakami}, \citenamefont {Bostwick},
  \citenamefont {Rotenberg},\ and\ \citenamefont {Jozwiak}}]{Katoch:2018}%
  \BibitemOpen
  \bibfield  {author} {\bibinfo {author} {\bibfnamefont {J.}~\bibnamefont
  {Katoch}}, \bibinfo {author} {\bibfnamefont {S.}~\bibnamefont {Ulstrup}},
  \bibinfo {author} {\bibfnamefont {R.~J.}\ \bibnamefont {Koch}}, \bibinfo
  {author} {\bibfnamefont {S.}~\bibnamefont {Moser}}, \bibinfo {author}
  {\bibfnamefont {K.}~\bibnamefont {Mc{C}reary}}, \bibinfo {author}
  {\bibfnamefont {S.}~\bibnamefont {Singh}}, \bibinfo {author} {\bibfnamefont
  {J.}~\bibnamefont {Xu}}, \bibinfo {author} {\bibfnamefont {B.}~\bibnamefont
  {Jonker}}, \bibinfo {author} {\bibfnamefont {R.}~\bibnamefont {Kawakami}},
  \bibinfo {author} {\bibfnamefont {A.}~\bibnamefont {Bostwick}}, \bibinfo
  {author} {\bibfnamefont {E.}~\bibnamefont {Rotenberg}}, \ and\ \bibinfo
  {author} {\bibfnamefont {C.}~\bibnamefont {Jozwiak}},\ }\href@noop {}
  {\bibfield  {journal} {\bibinfo  {journal} {Nature Physics}\ }\textbf
  {\bibinfo {volume} {14}},\ \bibinfo {pages} {355} (\bibinfo {year}
  {2018})}\BibitemShut {NoStop}%
\bibitem [{\citenamefont {Chen}\ \emph {et~al.}(2008)\citenamefont {Chen},
  \citenamefont {Jang}, \citenamefont {Xiao}, \citenamefont {Ishigami},\ and\
  \citenamefont {Fuhrer}}]{Chen:2008}%
  \BibitemOpen
  \bibfield  {author} {\bibinfo {author} {\bibfnamefont {J.-H.}\ \bibnamefont
  {Chen}}, \bibinfo {author} {\bibfnamefont {C.}~\bibnamefont {Jang}}, \bibinfo
  {author} {\bibfnamefont {S.}~\bibnamefont {Xiao}}, \bibinfo {author}
  {\bibfnamefont {M.}~\bibnamefont {Ishigami}}, \ and\ \bibinfo {author}
  {\bibfnamefont {M.~S.}\ \bibnamefont {Fuhrer}},\ }\href@noop {} {\bibfield
  {journal} {\bibinfo  {journal} {Nature Nanotechnology}\ }\textbf {\bibinfo
  {volume} {3}},\ \bibinfo {pages} {206} (\bibinfo {year} {2008})}\BibitemShut
  {NoStop}%
\bibitem [{\citenamefont {Dean}\ \emph {et~al.}(2010)\citenamefont {Dean},
  \citenamefont {Young}, \citenamefont {Meric}, \citenamefont {Lee},
  \citenamefont {Wang}, \citenamefont {Sorgenfrei}, \citenamefont {Watanabe},
  \citenamefont {Taniguchi}, \citenamefont {Kim}, \citenamefont {Shepard},\
  and\ \citenamefont {Hone}}]{Dean:2010}%
  \BibitemOpen
  \bibfield  {author} {\bibinfo {author} {\bibfnamefont {C.}~\bibnamefont
  {Dean}}, \bibinfo {author} {\bibfnamefont {A.}~\bibnamefont {Young}},
  \bibinfo {author} {\bibfnamefont {I.}~\bibnamefont {Meric}}, \bibinfo
  {author} {\bibfnamefont {C.}~\bibnamefont {Lee}}, \bibinfo {author}
  {\bibfnamefont {L.}~\bibnamefont {Wang}}, \bibinfo {author} {\bibfnamefont
  {S.}~\bibnamefont {Sorgenfrei}}, \bibinfo {author} {\bibfnamefont
  {K.}~\bibnamefont {Watanabe}}, \bibinfo {author} {\bibfnamefont
  {T.}~\bibnamefont {Taniguchi}}, \bibinfo {author} {\bibfnamefont
  {P.}~\bibnamefont {Kim}}, \bibinfo {author} {\bibfnamefont {K.}~\bibnamefont
  {Shepard}}, \ and\ \bibinfo {author} {\bibfnamefont {J.}~\bibnamefont
  {Hone}},\ }\href {\doibase 10.1038/nnano.2010.172} {\bibfield  {journal}
  {\bibinfo  {journal} {Nature Nanotechnology}\ }\textbf {\bibinfo {volume}
  {5}},\ \bibinfo {pages} {722} (\bibinfo {year} {2010})}\BibitemShut {NoStop}%
\bibitem [{\citenamefont {R\"osner}\ \emph {et~al.}(2016)\citenamefont
  {R\"osner}, \citenamefont {Steinke}, \citenamefont {Lorke}, \citenamefont
  {Gies}, \citenamefont {Jahnke},\ and\ \citenamefont {Wehling}}]{Rosner:2016}%
  \BibitemOpen
  \bibfield  {author} {\bibinfo {author} {\bibfnamefont {M.}~\bibnamefont
  {R\"osner}}, \bibinfo {author} {\bibfnamefont {C.}~\bibnamefont {Steinke}},
  \bibinfo {author} {\bibfnamefont {M.}~\bibnamefont {Lorke}}, \bibinfo
  {author} {\bibfnamefont {C.}~\bibnamefont {Gies}}, \bibinfo {author}
  {\bibfnamefont {F.}~\bibnamefont {Jahnke}}, \ and\ \bibinfo {author}
  {\bibfnamefont {T.~O.}\ \bibnamefont {Wehling}},\ }\href {\doibase
  10.1021/acs.nanolett.5b05009} {\bibfield  {journal} {\bibinfo  {journal}
  {Nano Letters}\ }\textbf {\bibinfo {volume} {16}},\ \bibinfo {pages} {2322}
  (\bibinfo {year} {2016})}\BibitemShut {NoStop}%
\bibitem [{\citenamefont {Raja}\ \emph {et~al.}(2017)\citenamefont {Raja},
  \citenamefont {Chaves}, \citenamefont {Yu}, \citenamefont {Arefe},
  \citenamefont {Hill}, \citenamefont {Rigosi}, \citenamefont {Berkelbach},
  \citenamefont {Nagler}, \citenamefont {Sch{\"u}ller}, \citenamefont {Korn},
  \citenamefont {Nuckolls}, \citenamefont {Hone}, \citenamefont {Brus},
  \citenamefont {Heinz}, \citenamefont {Reichman},\ and\ \citenamefont
  {Chernikov}}]{Raja:2017}%
  \BibitemOpen
  \bibfield  {author} {\bibinfo {author} {\bibfnamefont {A.}~\bibnamefont
  {Raja}}, \bibinfo {author} {\bibfnamefont {A.}~\bibnamefont {Chaves}},
  \bibinfo {author} {\bibfnamefont {J.}~\bibnamefont {Yu}}, \bibinfo {author}
  {\bibfnamefont {G.}~\bibnamefont {Arefe}}, \bibinfo {author} {\bibfnamefont
  {H.~M.}\ \bibnamefont {Hill}}, \bibinfo {author} {\bibfnamefont {A.~F.}\
  \bibnamefont {Rigosi}}, \bibinfo {author} {\bibfnamefont {T.~C.}\
  \bibnamefont {Berkelbach}}, \bibinfo {author} {\bibfnamefont
  {P.}~\bibnamefont {Nagler}}, \bibinfo {author} {\bibfnamefont
  {C.}~\bibnamefont {Sch{\"u}ller}}, \bibinfo {author} {\bibfnamefont
  {T.}~\bibnamefont {Korn}}, \bibinfo {author} {\bibfnamefont {C.}~\bibnamefont
  {Nuckolls}}, \bibinfo {author} {\bibfnamefont {J.}~\bibnamefont {Hone}},
  \bibinfo {author} {\bibfnamefont {L.~E.}\ \bibnamefont {Brus}}, \bibinfo
  {author} {\bibfnamefont {T.~F.}\ \bibnamefont {Heinz}}, \bibinfo {author}
  {\bibfnamefont {D.~R.}\ \bibnamefont {Reichman}}, \ and\ \bibinfo {author}
  {\bibfnamefont {A.}~\bibnamefont {Chernikov}},\ }\href
  {https://doi.org/10.1038/ncomms15251} {\bibfield  {journal} {\bibinfo
  {journal} {Nature Communications}\ }\textbf {\bibinfo {volume} {8}},\
  \bibinfo {pages} {15251} (\bibinfo {year} {2017})}\BibitemShut {NoStop}%
\bibitem [{\citenamefont {Ghatak}\ \emph {et~al.}(2011)\citenamefont {Ghatak},
  \citenamefont {Pal},\ and\ \citenamefont {Ghosh}}]{Ghatak:2011}%
  \BibitemOpen
  \bibfield  {author} {\bibinfo {author} {\bibfnamefont {S.}~\bibnamefont
  {Ghatak}}, \bibinfo {author} {\bibfnamefont {A.~N.}\ \bibnamefont {Pal}}, \
  and\ \bibinfo {author} {\bibfnamefont {A.}~\bibnamefont {Ghosh}},\ }\href
  {\doibase 10.1021/nn202852j} {\bibfield  {journal} {\bibinfo  {journal} {ACS
  Nano}\ }\textbf {\bibinfo {volume} {5}},\ \bibinfo {pages} {7707} (\bibinfo
  {year} {2011})}\BibitemShut {NoStop}%
\bibitem [{\citenamefont {Khestanova}\ \emph {et~al.}(2016)\citenamefont
  {Khestanova}, \citenamefont {Guinea}, \citenamefont {Fumagalli},
  \citenamefont {Geim},\ and\ \citenamefont {Grigorieva}}]{Khestanova:2016}%
  \BibitemOpen
  \bibfield  {author} {\bibinfo {author} {\bibfnamefont {E.}~\bibnamefont
  {Khestanova}}, \bibinfo {author} {\bibfnamefont {F.}~\bibnamefont {Guinea}},
  \bibinfo {author} {\bibfnamefont {L.}~\bibnamefont {Fumagalli}}, \bibinfo
  {author} {\bibfnamefont {A.~K.}\ \bibnamefont {Geim}}, \ and\ \bibinfo
  {author} {\bibfnamefont {I.~V.}\ \bibnamefont {Grigorieva}},\ }\href
  {https://doi.org/10.1038/ncomms12587} {\bibfield  {journal} {\bibinfo
  {journal} {Nature Communications}\ }\textbf {\bibinfo {volume} {7}},\
  \bibinfo {pages} {12587} (\bibinfo {year} {2016})}\BibitemShut {NoStop}%
\bibitem [{\citenamefont {Klein}\ \emph {et~al.}(2001)\citenamefont {Klein},
  \citenamefont {Tiefenbacher}, \citenamefont {Eyert}, \citenamefont
  {Pettenkofer},\ and\ \citenamefont {Jaegermann}}]{Klein2001}%
  \BibitemOpen
  \bibfield  {author} {\bibinfo {author} {\bibfnamefont {A.}~\bibnamefont
  {Klein}}, \bibinfo {author} {\bibfnamefont {S.}~\bibnamefont {Tiefenbacher}},
  \bibinfo {author} {\bibfnamefont {V.}~\bibnamefont {Eyert}}, \bibinfo
  {author} {\bibfnamefont {C.}~\bibnamefont {Pettenkofer}}, \ and\ \bibinfo
  {author} {\bibfnamefont {W.}~\bibnamefont {Jaegermann}},\ }\href@noop {}
  {\bibfield  {journal} {\bibinfo  {journal} {Phys. Rev. B}\ }\textbf {\bibinfo
  {volume} {64}} (\bibinfo {year} {2001})}\BibitemShut {NoStop}%
\bibitem [{\citenamefont {Yuan}\ \emph {et~al.}(2016)\citenamefont {Yuan},
  \citenamefont {Liu}, \citenamefont {Xu}, \citenamefont {Zhou}, \citenamefont
  {Wu}, \citenamefont {Dumcenco}, \citenamefont {Yan}, \citenamefont {Zhang},
  \citenamefont {Mo}, \citenamefont {Dudin}, \citenamefont {Kandyba},
  \citenamefont {Yablonskikh}, \citenamefont {Barinov}, \citenamefont {Shen},
  \citenamefont {Zhang}, \citenamefont {Huang}, \citenamefont {Xu},
  \citenamefont {Hussain}, \citenamefont {Hwang}, \citenamefont {Cui},\ and\
  \citenamefont {Chen}}]{Yuan:2016}%
  \BibitemOpen
  \bibfield  {author} {\bibinfo {author} {\bibfnamefont {H.}~\bibnamefont
  {Yuan}}, \bibinfo {author} {\bibfnamefont {Z.}~\bibnamefont {Liu}}, \bibinfo
  {author} {\bibfnamefont {G.}~\bibnamefont {Xu}}, \bibinfo {author}
  {\bibfnamefont {B.}~\bibnamefont {Zhou}}, \bibinfo {author} {\bibfnamefont
  {S.}~\bibnamefont {Wu}}, \bibinfo {author} {\bibfnamefont {D.}~\bibnamefont
  {Dumcenco}}, \bibinfo {author} {\bibfnamefont {K.}~\bibnamefont {Yan}},
  \bibinfo {author} {\bibfnamefont {Y.}~\bibnamefont {Zhang}}, \bibinfo
  {author} {\bibfnamefont {S.-K.}\ \bibnamefont {Mo}}, \bibinfo {author}
  {\bibfnamefont {P.}~\bibnamefont {Dudin}}, \bibinfo {author} {\bibfnamefont
  {V.}~\bibnamefont {Kandyba}}, \bibinfo {author} {\bibfnamefont
  {M.}~\bibnamefont {Yablonskikh}}, \bibinfo {author} {\bibfnamefont
  {A.}~\bibnamefont {Barinov}}, \bibinfo {author} {\bibfnamefont
  {Z.}~\bibnamefont {Shen}}, \bibinfo {author} {\bibfnamefont {S.}~\bibnamefont
  {Zhang}}, \bibinfo {author} {\bibfnamefont {Y.}~\bibnamefont {Huang}},
  \bibinfo {author} {\bibfnamefont {X.}~\bibnamefont {Xu}}, \bibinfo {author}
  {\bibfnamefont {Z.}~\bibnamefont {Hussain}}, \bibinfo {author} {\bibfnamefont
  {H.~Y.}\ \bibnamefont {Hwang}}, \bibinfo {author} {\bibfnamefont
  {Y.}~\bibnamefont {Cui}}, \ and\ \bibinfo {author} {\bibfnamefont
  {Y.}~\bibnamefont {Chen}},\ }\href {\doibase 10.1021/acs.nanolett.5b05107}
  {\bibfield  {journal} {\bibinfo  {journal} {Nano Letters}\ }\textbf {\bibinfo
  {volume} {16}},\ \bibinfo {pages} {4738} (\bibinfo {year}
  {2016})}\BibitemShut {NoStop}%
\bibitem [{\citenamefont {Lui}\ \emph {et~al.}(2009)\citenamefont {Lui},
  \citenamefont {Liu}, \citenamefont {Mak}, \citenamefont {Flynn},\ and\
  \citenamefont {Heinz}}]{lui2009}%
  \BibitemOpen
  \bibfield  {author} {\bibinfo {author} {\bibfnamefont {C.~H.}\ \bibnamefont
  {Lui}}, \bibinfo {author} {\bibfnamefont {L.}~\bibnamefont {Liu}}, \bibinfo
  {author} {\bibfnamefont {K.~F.}\ \bibnamefont {Mak}}, \bibinfo {author}
  {\bibfnamefont {G.~W.}\ \bibnamefont {Flynn}}, \ and\ \bibinfo {author}
  {\bibfnamefont {T.~F.}\ \bibnamefont {Heinz}},\ }\href@noop {} {\bibfield
  {journal} {\bibinfo  {journal} {Nature}\ }\textbf {\bibinfo {volume} {462}},\
  \bibinfo {pages} {339} (\bibinfo {year} {2009})}\BibitemShut {NoStop}%
\bibitem [{\citenamefont {Chernikov}\ \emph {et~al.}(2015)\citenamefont
  {Chernikov}, \citenamefont {Zande}, \citenamefont {Hill}, \citenamefont
  {Rigosi}, \citenamefont {Velauthapillai}, \citenamefont {Hone},\ and\
  \citenamefont {Heinz}}]{Chernikov:2015b}%
  \BibitemOpen
  \bibfield  {author} {\bibinfo {author} {\bibfnamefont {A.}~\bibnamefont
  {Chernikov}}, \bibinfo {author} {\bibfnamefont {A.}~\bibnamefont {Zande}},
  \bibinfo {author} {\bibfnamefont {H.}~\bibnamefont {Hill}}, \bibinfo {author}
  {\bibfnamefont {A.}~\bibnamefont {Rigosi}}, \bibinfo {author} {\bibfnamefont
  {A.}~\bibnamefont {Velauthapillai}}, \bibinfo {author} {\bibfnamefont
  {J.}~\bibnamefont {Hone}}, \ and\ \bibinfo {author} {\bibfnamefont
  {T.}~\bibnamefont {Heinz}},\ }\href {\doibase 10.1103/PhysRevLett.115.126802}
  {\bibfield  {journal} {\bibinfo  {journal} {Physical Review Letters}\
  }\textbf {\bibinfo {volume} {115}},\ \bibinfo {pages} {126802} (\bibinfo
  {year} {2015})}\BibitemShut {NoStop}%
\bibitem [{\citenamefont {Kresse}\ \emph {et~al.}(2012)\citenamefont {Kresse},
  \citenamefont {Marsman}, \citenamefont {Hintzsche},\ and\ \citenamefont
  {Flage-Larsen}}]{Kresse:2012}%
  \BibitemOpen
  \bibfield  {author} {\bibinfo {author} {\bibfnamefont {G.}~\bibnamefont
  {Kresse}}, \bibinfo {author} {\bibfnamefont {M.}~\bibnamefont {Marsman}},
  \bibinfo {author} {\bibfnamefont {L.~E.}\ \bibnamefont {Hintzsche}}, \ and\
  \bibinfo {author} {\bibfnamefont {E.}~\bibnamefont {Flage-Larsen}},\ }\href
  {\doibase 10.1103/PhysRevB.85.045205} {\bibfield  {journal} {\bibinfo
  {journal} {Phys. Rev. B}\ }\textbf {\bibinfo {volume} {85}},\ \bibinfo
  {pages} {045205} (\bibinfo {year} {2012})}\BibitemShut {NoStop}%
\bibitem [{\citenamefont {Williams}(1965)}]{Williams:1965}%
  \BibitemOpen
  \bibfield  {author} {\bibinfo {author} {\bibfnamefont {R.}~\bibnamefont
  {Williams}},\ }\href {\doibase 10.1103/PhysRev.140.A569} {\bibfield
  {journal} {\bibinfo  {journal} {Phys. Rev.}\ }\textbf {\bibinfo {volume}
  {140}},\ \bibinfo {pages} {A569} (\bibinfo {year} {1965})}\BibitemShut
  {NoStop}%
\bibitem [{\citenamefont {Landmann}\ \emph {et~al.}(2012)\citenamefont
  {Landmann}, \citenamefont {Rauls},\ and\ \citenamefont
  {Schmidt}}]{Landmann:2012}%
  \BibitemOpen
  \bibfield  {author} {\bibinfo {author} {\bibfnamefont {M.}~\bibnamefont
  {Landmann}}, \bibinfo {author} {\bibfnamefont {E.}~\bibnamefont {Rauls}}, \
  and\ \bibinfo {author} {\bibfnamefont {W.~G.}\ \bibnamefont {Schmidt}},\
  }\href {http://stacks.iop.org/0953-8984/24/i=19/a=195503} {\bibfield
  {journal} {\bibinfo  {journal} {Journal of Physics: Condensed Matter}\
  }\textbf {\bibinfo {volume} {24}},\ \bibinfo {pages} {195503} (\bibinfo
  {year} {2012})}\BibitemShut {NoStop}%
\bibitem [{\citenamefont {Scanlon}\ \emph {et~al.}(2013)\citenamefont
  {Scanlon}, \citenamefont {Dunnill}, \citenamefont {Buckeridge}, \citenamefont
  {Shevlin}, \citenamefont {Logsdail}, \citenamefont {Woodley}, \citenamefont
  {Catlow}, \citenamefont {Powell}, \citenamefont {Palgrave}, \citenamefont
  {Parkin}, \citenamefont {Watson}, \citenamefont {Keal}, \citenamefont
  {Sherwood}, \citenamefont {Walsh},\ and\ \citenamefont
  {Sokol}}]{Scanlon:2013}%
  \BibitemOpen
  \bibfield  {author} {\bibinfo {author} {\bibfnamefont {D.~O.}\ \bibnamefont
  {Scanlon}}, \bibinfo {author} {\bibfnamefont {C.~W.}\ \bibnamefont
  {Dunnill}}, \bibinfo {author} {\bibfnamefont {J.}~\bibnamefont {Buckeridge}},
  \bibinfo {author} {\bibfnamefont {S.~A.}\ \bibnamefont {Shevlin}}, \bibinfo
  {author} {\bibfnamefont {A.~J.}\ \bibnamefont {Logsdail}}, \bibinfo {author}
  {\bibfnamefont {S.~M.}\ \bibnamefont {Woodley}}, \bibinfo {author}
  {\bibfnamefont {C.~R.~A.}\ \bibnamefont {Catlow}}, \bibinfo {author}
  {\bibfnamefont {M.~J.}\ \bibnamefont {Powell}}, \bibinfo {author}
  {\bibfnamefont {R.~G.}\ \bibnamefont {Palgrave}}, \bibinfo {author}
  {\bibfnamefont {I.~P.}\ \bibnamefont {Parkin}}, \bibinfo {author}
  {\bibfnamefont {G.~W.}\ \bibnamefont {Watson}}, \bibinfo {author}
  {\bibfnamefont {T.~W.}\ \bibnamefont {Keal}}, \bibinfo {author}
  {\bibfnamefont {P.}~\bibnamefont {Sherwood}}, \bibinfo {author}
  {\bibfnamefont {A.}~\bibnamefont {Walsh}}, \ and\ \bibinfo {author}
  {\bibfnamefont {A.~A.}\ \bibnamefont {Sokol}},\ }\href
  {https://doi.org/10.1038/nmat3697} {\bibfield  {journal} {\bibinfo  {journal}
  {Nature Materials}\ }\textbf {\bibinfo {volume} {12}},\ \bibinfo {pages} {798} (\bibinfo {year} {2013})}\BibitemShut {NoStop}%
\bibitem [{\citenamefont {Arnaud}\ \emph {et~al.}(2006)\citenamefont {Arnaud},
  \citenamefont {Leb\`egue}, \citenamefont {Rabiller},\ and\ \citenamefont
  {Alouani}}]{Arnaud:2006}%
  \BibitemOpen
  \bibfield  {author} {\bibinfo {author} {\bibfnamefont {B.}~\bibnamefont
  {Arnaud}}, \bibinfo {author} {\bibfnamefont {S.}~\bibnamefont {Leb\`egue}},
  \bibinfo {author} {\bibfnamefont {P.}~\bibnamefont {Rabiller}}, \ and\
  \bibinfo {author} {\bibfnamefont {M.}~\bibnamefont {Alouani}},\ }\href
  {\doibase 10.1103/PhysRevLett.96.026402} {\bibfield  {journal} {\bibinfo
  {journal} {Phys. Rev. Lett.}\ }\textbf {\bibinfo {volume} {96}},\ \bibinfo
  {pages} {026402} (\bibinfo {year} {2006})}\BibitemShut {NoStop}%
\bibitem [{\citenamefont {Sup~Choi}\ \emph {et~al.}(2013)\citenamefont
  {Sup~Choi}, \citenamefont {Lee}, \citenamefont {Yu}, \citenamefont {Lee},
  \citenamefont {Hwan~Lee}, \citenamefont {Kim}, \citenamefont {Hone},\ and\
  \citenamefont {Jong~Yoo}}]{Choi:2013}%
  \BibitemOpen
  \bibfield  {author} {\bibinfo {author} {\bibfnamefont {M.}~\bibnamefont
  {Sup~Choi}}, \bibinfo {author} {\bibfnamefont {G.-H.}\ \bibnamefont {Lee}},
  \bibinfo {author} {\bibfnamefont {Y.-J.}\ \bibnamefont {Yu}}, \bibinfo
  {author} {\bibfnamefont {D.-Y.}\ \bibnamefont {Lee}}, \bibinfo {author}
  {\bibfnamefont {S.}~\bibnamefont {Hwan~Lee}}, \bibinfo {author}
  {\bibfnamefont {P.}~\bibnamefont {Kim}}, \bibinfo {author} {\bibfnamefont
  {J.}~\bibnamefont {Hone}}, \ and\ \bibinfo {author} {\bibfnamefont
  {W.}~\bibnamefont {Jong~Yoo}},\ }\href {https://doi.org/10.1038/ncomms2652}
  {\bibfield  {journal} {\bibinfo  {journal} {Nature Communications}\ }\textbf
  {\bibinfo {volume} {4}},\ \bibinfo {pages} {1624} (\bibinfo {year}
  {2013})}\BibitemShut {NoStop}%
\bibitem [{\citenamefont {Fiori}\ \emph {et~al.}(2012)\citenamefont {Fiori},
  \citenamefont {Betti}, \citenamefont {Bruzzone},\ and\ \citenamefont
  {Iannaccone}}]{Fiori:2012}%
  \BibitemOpen
  \bibfield  {author} {\bibinfo {author} {\bibfnamefont {G.}~\bibnamefont
  {Fiori}}, \bibinfo {author} {\bibfnamefont {A.}~\bibnamefont {Betti}},
  \bibinfo {author} {\bibfnamefont {S.}~\bibnamefont {Bruzzone}}, \ and\
  \bibinfo {author} {\bibfnamefont {G.}~\bibnamefont {Iannaccone}},\ }\href
  {\doibase 10.1021/nn300019b} {\bibfield  {journal} {\bibinfo  {journal} {ACS
  Nano}\ }\textbf {\bibinfo {volume} {6}},\ \bibinfo {pages} {2642} (\bibinfo
  {year} {2012})}\BibitemShut {NoStop}%
\bibitem [{\citenamefont {Powers}\ \emph {et~al.}(1995)\citenamefont {Powers},
  \citenamefont {Benjamin}, \citenamefont {Porter}, \citenamefont {Nemanich},
  \citenamefont {Davis}, \citenamefont {Cuomo}, \citenamefont {Doll},\ and\
  \citenamefont {Harris}}]{Powers:1995}%
  \BibitemOpen
  \bibfield  {author} {\bibinfo {author} {\bibfnamefont {M.~J.}\ \bibnamefont
  {Powers}}, \bibinfo {author} {\bibfnamefont {M.~C.}\ \bibnamefont
  {Benjamin}}, \bibinfo {author} {\bibfnamefont {L.~M.}\ \bibnamefont
  {Porter}}, \bibinfo {author} {\bibfnamefont {R.~J.}\ \bibnamefont
  {Nemanich}}, \bibinfo {author} {\bibfnamefont {R.~F.}\ \bibnamefont {Davis}},
  \bibinfo {author} {\bibfnamefont {J.~J.}\ \bibnamefont {Cuomo}}, \bibinfo
  {author} {\bibfnamefont {G.~L.}\ \bibnamefont {Doll}}, \ and\ \bibinfo
  {author} {\bibfnamefont {S.~J.}\ \bibnamefont {Harris}},\ }\href {\doibase
  10.1063/1.115315} {\bibfield  {journal} {\bibinfo  {journal} {Applied Physics
  Letters}\ }\textbf {\bibinfo {volume} {67}},\ \bibinfo {pages} {3912}
  (\bibinfo {year} {1995})}\BibitemShut {NoStop}%
\bibitem [{\citenamefont {Onda}\ \emph {et~al.}(2004)\citenamefont {Onda},
  \citenamefont {Li},\ and\ \citenamefont {Petek}}]{Onda2004}%
  \BibitemOpen
  \bibfield  {author} {\bibinfo {author} {\bibfnamefont {K.}~\bibnamefont
  {Onda}}, \bibinfo {author} {\bibfnamefont {B.}~\bibnamefont {Li}}, \ and\
  \bibinfo {author} {\bibfnamefont {H.}~\bibnamefont {Petek}},\ }\href@noop {}
  {\bibfield  {journal} {\bibinfo  {journal} {Physical Review B}\ }\textbf
  {\bibinfo {volume} {70}},\ \bibinfo {pages} {045415} (\bibinfo {year}
  {2004})}\BibitemShut {NoStop}%
\bibitem [{\citenamefont {Kaminski}\ \emph {et~al.}(2016)\citenamefont
  {Kaminski}, \citenamefont {Rosenkranz}, \citenamefont {Norman}, \citenamefont
  {Randeria}, \citenamefont {Li}, \citenamefont {Raffy},\ and\ \citenamefont
  {Campuzano}}]{Kaminski:2016}%
  \BibitemOpen
  \bibfield  {author} {\bibinfo {author} {\bibfnamefont {A.}~\bibnamefont
  {Kaminski}}, \bibinfo {author} {\bibfnamefont {S.}~\bibnamefont
  {Rosenkranz}}, \bibinfo {author} {\bibfnamefont {M.~R.}\ \bibnamefont
  {Norman}}, \bibinfo {author} {\bibfnamefont {M.}~\bibnamefont {Randeria}},
  \bibinfo {author} {\bibfnamefont {Z.~Z.}\ \bibnamefont {Li}}, \bibinfo
  {author} {\bibfnamefont {H.}~\bibnamefont {Raffy}}, \ and\ \bibinfo {author}
  {\bibfnamefont {J.~C.}\ \bibnamefont {Campuzano}},\ }\href {\doibase
  10.1103/PhysRevX.6.031040} {\bibfield  {journal} {\bibinfo  {journal} {Phys.
  Rev. X}\ }\textbf {\bibinfo {volume} {6}},\ \bibinfo {pages} {031040}
  (\bibinfo {year} {2016})}\BibitemShut {NoStop}%
\end{thebibliography}
\end{document}